\documentclass[conference]{IEEEtran}

\IEEEoverridecommandlockouts

\usepackage{cite}
\usepackage{amsmath,amssymb,amsfonts}
\usepackage{relsize}
\usepackage{algorithmic}
\usepackage{graphicx}
\usepackage{textcomp}
\usepackage{xcolor}
\usepackage{hyperref}

\usepackage{amsmath}
\DeclareMathOperator*{\argmax}{argmax\,}
\DeclareMathOperator*{\argmin}{argmin\,}
\newcommand\norm[1]{\lVert#1\rVert}

\begin{document}

\title{Score-Based Generative Models for Robust Channel Estimation
\thanks{This work was supported by ONR grant N00014-19-1-2590.}
}

\author{\IEEEauthorblockN{Marius Arvinte}
\IEEEauthorblockA{\textit{Electrical and Computer Engineering} \\
\textit{University of Texas at Austin}\\
Austin, TX, USA \\
arvinte@utexas.edu}
\and
\IEEEauthorblockN{Jonathan I. Tamir}
\IEEEauthorblockA{\textit{Electrical and Computer Engineering} \\
\textit{University of Texas at Austin}\\
Austin, TX, USA \\
jtamir@utexas.edu}
}

\maketitle
\begin{abstract}
Channel estimation is a critical task in digital communications that greatly impacts end-to-end system performance. In this work, we introduce a novel approach for multiple-input multiple-output (MIMO) channel estimation using \textit{score-based generative models}. Our method uses a deep neural network that is trained to estimate the gradient of the log-prior of wireless channels at any point in high-dimensional space, and leverages this model to solve channel estimation via \textit{posterior sampling}. We train a score-based model on channel realizations from the CDL-D model for two antenna spacings and show that the approach leads to competitive in- and out-of-distribution performance when compared to generative adversarial network (GAN) and compressed sensing (CS) methods. When tested on CDL-D channels, the approach leads to a gain of at least $5$ dB in channel estimation error compared to GAN methods in-distribution at $\lambda/2$ antenna spacing. When tested on CDL-C channels which are never seen during training or fine-tuned on, the approach leads to end-to-end coded performance gains of up to $3$ dB compared to CS methods and losses of only $0.5$ dB compared to ideal channel knowledge.
\end{abstract}

\section{Introduction}
Channel estimation represents the task of recovering a communications channel tensor using a number of measurements. In the vast majority of cases, these measurements come from pilot symbols transmitted at specific intervals, which are known ahead of time by the receiver. This task is critical for the end-to-end performance, since the channel estimate is used in downstream tasks, such as symbol equalization and soft demodulation of the data transmissions, and is thus an active area of research. With the adoption of 5G and development of 6G standards \cite{heng2021six}, channel estimation via deep learning has been identified as a promising solution, especially in high-dimensional cases such as MIMO millimeter-wave communications \cite{heath2016overview,balevi2020high}.

Our main motivation in this work is to develop robust, data-driven MIMO channel estimation algorithms: we want our method to work reasonably well under distributional shifts compared to the training conditions \emph{without} adaptation. Although online adaptation methods are possible \cite{liu2019online}, they still require a set of ground-truth channels from the test distribution to be known, which may incur a significant pilot overhead. Thus, algorithms that generalize and retain their performance in completely novel test-time distributions without fine-tuning are an important component toward real-world utilization. To exemplify robustness in the results section, we train our method on CDL-D line of sight channels, and test it on CDL-C non-line of sight channels, and show that the proposed method retains its performance.

In this paper, we use score-based generative models \cite{song2019generative} for robust, high-dimensional channel estimation. Inspired by recent research that shows the potential of this method for magnetic resonance imaging (MRI) reconstruction \cite{jalal2021robust}, we introduce a training and hyper-parameter tuning approach for wireless channel estimation in a single-carrier MIMO communication scenario. For training, a database of known channels is used to train a score-based model in an unsupervised manner that is independent of the pilot symbols. During inference, we tackle the main challenges that arise when applying this method to channel estimation: the need to operate in a very wide (e.g., up to $30$ dB) signal-to-noise ratio (SNR) range, as well as robust out-of-distribution performance. Our experiments show that with the proposed choice of hyper-parameters, test-time performance is retained even in simulated test channels that come from a completely different distribution than during training, and gains of up to $3$ dB are achieved against compressed sensing based methods in end-to-end system simulations. A complete implementation of our algorithm is available online\footnote{\href{https://github.com/utcsilab/diffusion-channels}{\texttt{https://github.com/utcsilab/diffusion-channels}}}.

\subsection{Related Work}
Previous work on data-driven deep learning methods for channel estimation largely falls in the category of supervised methods. In \cite{soltani2019deep}, the matrix of pilots (with time and frequency dimensions) is passed through a deep convolutional neural network that is trained to output the ground-truth channel that generates the pilots. The work in \cite{hu2020deep} follows the same approach and shows that outside of the training distribution, a severe performance loss (tens of dB) occurs. This is not the case for score-based models, which are shown to be theoretically robust outside of the training set under certain assumptions, as well as in practice \cite{jalal2021robust}. The work in \cite{balevi2020high} takes a different approach than supervised learning by applying the compressed sensing with generative models (CGSM) to channel estimation by training a deep generative adversarial network to serve as an implicit prior. Their results show excellent performance in the low SNR and few-measurements regime, but saturated performance in the high-SNR or high-measurements regimes \cite{bora2017compressed}. This serves as a strong baseline for our approach, since both methods are unsupervised.

Score-based generative models are originally developed in \cite{song2019generative}, where the authors also introduce annealed Langevin dynamics to efficiently sample from a distribution of interest using a learned deep neural network. The work in \cite{song2020improved} introduces practical techniques for improving the training of score-based generative models, which we fully leverage. The work in \cite{JalalKDP21} theoretically proves the in-distribution optimality of posterior sampling and links this to their fully-dimensional support. Finally, the work in \cite{jalal2021robust} investigates the robustness of posterior sampling in the setting of MRI reconstruction, where it is shown that these models are competitive with state-of-the-art deep learning methods in-distribution and significantly more robust out-of-distribution.

\subsection{Contributions}
Our contributions in this work are the following:
\begin{enumerate}
    \item We formulate channel estimation via \textit{posterior sampling}: given measurements $\mathbf{y}$ obtained from pilot transmissions, we estimate the underlying channel $\mathbf{h}$ by sampling from the posterior distribution $p_H(\mathbf{h}|\mathbf{y})$.
    \item We successfully train a noise-conditional score-based generative model for millimeter-wave wireless channel matrices from the CDL-D channel model and show that this can be used to estimate channels even when a reduced number of pilots makes the inverse problem under-determined.
    \item We develop a hyper-parameter selection algorithm for posterior sampling with score-based models that optimizes the average relative channel estimation error over a wide range of operating SNR values. We test our scheme with and without knowledge of the exact operating SNR and show superior estimation performance both in- and \textit{out-of-distribution} (on CDL-C channels) with \textit{zero} online adaptation or fine-tuning.
\end{enumerate}

\section{Channel Estimation as an Inverse Problem}
We consider MIMO channel estimation in a narrowband digital communication scenario. Given a complex-valued channel matrix $\mathbf{H} \in \mathbb{C}^{N_\mathrm{r}\times N_\mathrm{t}}$ and the matrix of complex-valued pilots $\mathbf{P} \in \mathbb{C}^{N_\mathrm{t}\times N_\mathrm{p}}$ the measurements seen at the receiver are expressed as:
\begin{equation}
    \mathbf{Y} = \mathbf{H}\mathbf{P} + \mathbf{N},
    \label{eq:basic_model}
\end{equation}
\noindent where $\mathbf{N}$ is complex additive white Gaussian noise, $N_\mathrm{t}$ and $N_\mathrm{r}$ are the number of transmitter and receiver antennas, respectively, and $N_\mathrm{p}$ is the number of pilot transmissions.

In practice, the transmitter will form the matrix $\mathbf{P}$ as the product between a (potentially varying across pilots) beamforming matrix $\mathbf{W}_t$ and a matrix of pilot symbols $\mathbf{S}$. The beamforming matrix itself could be further factorized in its digital and analog parts \cite{heath2016overview}. In this work, we do not assume any kind of factorized structure for $\mathbf{P}$. This does not impact our method or the baselines, since we use unsupervised algorithms that are independent of the pilot matrix used. For the remainder of the paper, we also assume that the channel is constant across the pilot transmissions. In practice, this is a reasonable assumption if the pilots are staggered in time and the coherence time of the channel is much larger than the pilot sequence \cite{va2016impact}. The model in \eqref{eq:basic_model} is also of practical interest in OFDM scenarios, since it corresponds to a single pilot subcarrier.

Eq.~\eqref{eq:basic_model} can be rewritten by introducing the flattened channel and received symbol matrices $\mathbf{h}$ and $\mathbf{y}$, respectively, as:
\begin{equation}
    \mathbf{y} = \mathbf{A} \mathbf{h} + \mathbf{n}
    \label{eq:flat_model}
\end{equation}
\noindent where $\mathbf{A} \in \mathbb{C}^{N_\mathrm{p} N_\mathrm{r} \times N_\mathrm{t} N_\mathrm{r}}$ is the consolidated measurement matrix such that $\mathbf{A}= \mathbf{P}^T \otimes \mathbf{I}_{N_\mathrm{t}}$, and $\mathbf{n}$ is the flattened noise vector.

\begin{figure*}[!t]
\centering
\includegraphics[width=1\linewidth]{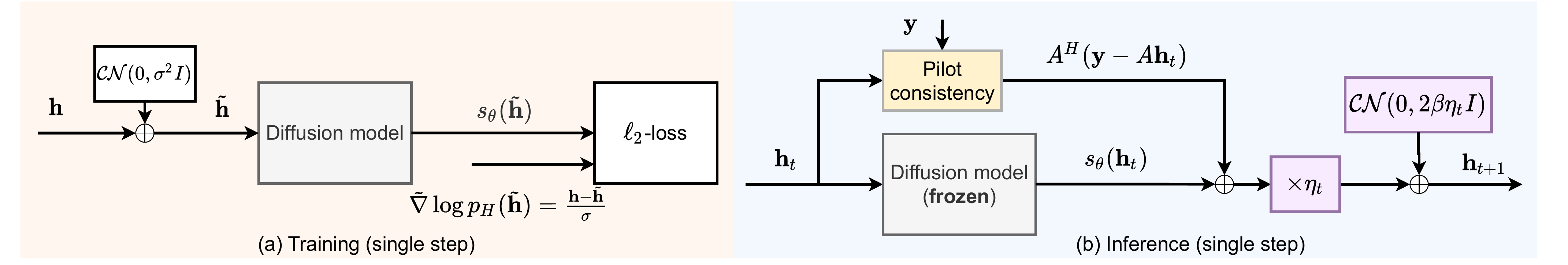}
\caption{(a) Block diagram of a training step. Noise is added to a training channel $\mathbf{h}$ and \eqref{eq:grad_train} is used to produce a regression target for the gradient of $\log p_H(\tilde{\mathbf{h}})$, after which a simple $\ell_2$-loss is used to train the parameters $\theta$ of the deep neural network via back-propagation. (b) Block diagram of a single inference step. The current channel estimate is updated by a pilot consistency term, a diffusion update, and added noise.}
\label{fig:train_infer}
\end{figure*}

Since channels of practical interest may severely deviate from the i.i.d. Gaussian assumption, especially in millimeter-wave scenarios \cite{heath2016overview}, the importance of a well-chosen prior when estimating $\mathbf{h}$ using the model in \eqref{eq:flat_model} is increased when the number of pilots makes the inverse problem under-determined. Denoting $\alpha = N_\mathrm{p} / N_\mathrm{t}$ as the undersampling ratio, a lower $\alpha$ leads to fewer pilots being used for channel estimation. When $\alpha < 1$, compressed sensing methods aim to estimate $\mathbf{h}$ by solving the following regularized optimization problem:
\begin{equation}
    \tilde{\mathbf{h}}_{\textrm{CS}} = \argmin_\mathbf{h} \norm{\mathbf{y}-\mathbf{A}\mathbf{h}}_2^2 + \lambda \mathcal{R}(\mathbf{h}),
    \label{eq:cs}
\end{equation}
\noindent where $\mathcal{R}$ is a sparsity-promoting regularization function and $\lambda \ge 0$ is a hyper-parameter that depends on the SNR and the chosen $\mathcal{R}$. When dealing with complicated signal distributions -- such as wireless channels -- this parameter is chosen in a data-driven way by using a set of validation channels and a search strategy (e.g., line search). A commonly used prior for wireless channel estimation is sparsity in the 2D Fourier (DFT) domain \cite{mendez2016hybrid}.

The approach of compressed sensing with generative models \cite{bora2017compressed,balevi2020high} has been applied to solve \eqref{eq:cs} by using deep generative adversarial networks as implicit priors. The algorithm recovers $\mathbf{h}$ by first training a GAN $G$ that takes as input a low-dimensional latent vector $\mathbf{z}$ and outputs a sample $\mathbf{h}$ from $p_H(\mathbf{h})$. With measurements $\mathbf{y}$, the channel is then estimated by solving the optimization problem:
\begin{multline}
    \tilde{\mathbf{h}}_{\textrm{GAN}} = G(\mathbf{z}^\star), \text{where} \\
    \mathbf{z}^\star = \argmin_\mathbf{z} \norm{\mathbf{y} - \mathbf{A}G(\mathbf{z})}_2^2 + \lambda_z \norm{\mathbf{z}}_2^2,
    \label{eq:csgm}
\end{multline}
\noindent and $\lambda_z$ enforces the Gaussian assumption placed on $\mathbf{z}$. To solve \eqref{eq:csgm} in practice, a gradient descent approach is used to find the solution $\mathbf{z}^\star$ since the deep neural network used is differentiable, and a line search is used for $\lambda_z$.

\section{Score-Based Generative Models for Channel Estimation}
\subsection{Score-Based Generative Models}
The key component of our approach is the class of models termed score-based generative models, which we briefly term score-based models, introduced in \cite{song2019generative}. These models are implicit prior models in that they are not trained to directly output $p_H(\mathbf{h}$. Instead, they are trained to estimate the gradient of the log-prior (score) of the underlying distribution $p_H(\mathbf{h})$ that governs a high-dimensional signal $\mathbf{h}$, such as MIMO channels. Given $\mathbf{h}$ sampled from $p_H(\mathbf{h})$, the goal of a score-based model $s_\theta$ with learnable weights $\theta$ is to output a close approximation of the gradient of the log-prior $\log p_H(\mathbf{h})$ with respect to $\mathbf{h}$:
\begin{equation}
    s_\theta(\mathbf{h}) \approx \nabla_\mathbf{h} \log p_H(\mathbf{h}).
\end{equation}

Since the exact gradients of $\log p_H(\mathbf{h})$ are intractable and unavailable for high-dimensional real-world signals, e.g., for wireless channels, an unsupervised approach is used to train score-based models with deep learning. To generate a target output for a given input $\mathbf{h}$, i.i.d. Gaussian noise $\mathbf{z}$ with zero mean and standard deviation $\sigma_z$ is added to $\mathbf{h}$, yielding a perturbed sample $\tilde{\mathbf{h}}$, for which the gradient with respect to $\tilde{\mathbf{h}}$ has the closed form:
\begin{equation}
    \tilde{\nabla}_\mathbf{\tilde{\mathbf{h}}} \log p_{\tilde{H}}(\tilde{\mathbf{h}}) = \frac{\mathbf{h} - \tilde{\mathbf{h}}}{\sigma_z^2} = -\frac{\mathbf{z}}{\sigma_z^2}.
    \label{eq:grad_train}
\end{equation}

Intuitively, the expression in \eqref{eq:grad_train} is a useful gradient since $\mathbf{z}/\sigma_z^2$ is the normalized direction that points away from the low density (perturbed, hence unnatural) point $\tilde{\mathbf{h}}$ to the high density one $\mathbf{h}$ in terms of the log-prior.

The approximate gradients are used to train the score-based model using the weighted mean squared error loss $L(\tilde{\mathbf{h}}; \theta, \sigma) = \norm{s_\theta(\tilde{\mathbf{h}}) - \sigma \tilde{\nabla}_\mathbf{h} \log p_H(\tilde{\mathbf{h}})}_2^2$. Importantly, during training, the value of $\sigma$ is randomly chosen from a broad interval at each training step, leading to score-based models learning gradients at a multitude of noise levels, including at points that fall outside of the support of $p_H(\mathbf{h})$.

This is a key aspect that aids the robustness of these models, since it effectively allows them to extrapolate and be well-behaved beyond the training dataset. In the sequel, we introduce channel estimation through posterior sampling, which is different than the previously considered approaches.

\subsection{Channel Estimation via Posterior Sampling}
Our proposed solution for channel estimation comes in the form of \textit{posterior sampling}. An overview of this approach, as well as a proof of optimality (given access to $p_H(\mathbf{h}|\mathbf{y})$) are given in \cite{JalalKDP21}. A block diagram of the entire approach is summarized in Fig.~\ref{fig:train_infer}. Fundamentally, we solve the estimation problem by sampling from the posterior, that is: $\tilde{\mathbf{h}} \sim p_H(\mathbf{h}|\mathbf{y})$.

In this work, we use annealed Langevin dynamics \cite{JalalKDP21} to sample from the posterior by running noisy gradient ascent, but other strategies are possible. At inference step $t+1$, the current channel estimate is updated using the rule:
\begin{equation}
    \mathbf{h}_{t+1} \leftarrow \mathbf{h}_t + \eta_t \nabla_{\mathbf{h}_t} \log p_H(\mathbf{h}_t | \mathbf{y}) + \sqrt{2 \beta \eta_t} \mathbf{\zeta}_t, \mathbf{\zeta}_t \sim \mathcal{CN}(0, I),
\end{equation}
\noindent where $\eta_t$ is the step size, $\beta$ is a hyper-parameter that controls the injected noise power, and $\mathbf{\zeta}_t$ is additive white Gaussian noise added at each step. Since the distribution $p_Y(\mathbf{y} | \textbf{h}_t)$ is Gaussian, we can apply Bayes rule to $p_H(\mathbf{h}_t | \mathbf{y})$ and omit the $p_Y(\mathbf{y})$ term since its gradient is zero with respect to $\mathbf{h}_t$. Furthermore, by using that the likelihood term can be written in closed form as: $\nabla_{\mathbf{h}_t} \log p_Y(\mathbf{y} | \mathbf{h}_t) = \mathbf{A}^H(\mathbf{A} \mathbf{h}_t - \mathbf{y})$, and that the deep score-based model satisfies $s_\theta(\mathbf{h}_t) \approx \nabla_{\mathbf{h}_t} \log p_H(\mathbf{h}_t)$, we obtain the final update rule as:
\begin{equation}
    \mathbf{h}_{t+1} \leftarrow \mathbf{h}_t + \eta_t s_\theta (\mathbf{h}_t) + \eta_t \mathbf{A}^H(\mathbf{y} - \mathbf{A} \mathbf{h}_t) + \sqrt{2 \beta \eta_t} \mathbf{\zeta}_t,
    \label{eq:final_update}
\end{equation}
\noindent where inference starts from a random $\mathbf{h}_0 \sim \mathcal{CN}(0, I)$ and is carried out for $N$ steps. Finally, the returned channel estimate is $\tilde{\mathbf{h}} = \mathbf{h}_N$.

The three terms involved in \eqref{eq:final_update} each contribute to solving the inverse the problem differently:
\begin{itemize}
    \item The first term takes a gradient step that maximizes the log-prior of the current channel estimate, i.e., makes the channel more naturally structured under its governing distribution $p_H(\mathbf{h})$.
    \item The second term takes a negative gradient step in the direction of the measurement loss, i.e., makes the channel more consistent with the observed pilots $\mathbf{y}$.
    \item The third term adds noise proportional to the step size at each step in order to prevent the approach from converging to the MAP estimate, which is sub-optimal when the posterior is multi-modal \cite{JalalKDP21}.
\end{itemize}

\section{Experimental Results}
\subsection{Training a Score-Based Model for MIMO Channels}
We train an NCSNv2 model \cite{song2020improved} on complex-valued channel matrices $\mathbf{H}$ with $N_\mathrm{t} = 64$ and $N_\mathrm{r} = 16$ as a proof-of-concept for our method. The backbone network we use is a RefineNet \cite{lin2017refinenet} with a total of $5.2$ million parameters and a depth of eight layers. To handle complex valued-inputs, we treat the real and imaginary parts of $\mathbf{H}$ as two input channels. The model is trained on a total of $20000$ training channel realizations drawn from the CDL-D channel model available in MATLAB, equally split between antenna spacings $\lambda/10$ and $\lambda/2$, each extracted from the first subcarrier of all symbols in a $14$ symbol frame, and using different random seeds.

Training for $800$ epochs takes approximately six hours on an NVIDIA RTX 2080Ti GPU. All estimation results come from a pilot matrix $\mathbf{P}$ containing random unit power QPSK symbols, but the methodology holds for any pilot sequence known ahead of time. To measure estimation quality we use the normalized mean square error metric, defined as $\textrm{NMSE} = \norm{\mathbf{H} - \tilde{\mathbf{H}}}_F^2 / \norm{\mathbf{H}}_F^2$. Complete details about the learning setup are available in the code.

\subsection{Tuning Inference Hyper-parameters}
We provide a method for tuning the hyper-parameters for the updates in \eqref{eq:final_update}. We do not tune the step size schedule (even though it could lead to reduced execution times) $\eta_t$ -- which is chosen according to \cite{song2020improved} -- but instead only tune two scalar hyper-parameters: the noise amplification factor $\beta$ and the number of iterations $N$. They both affect the final estimation performance: a $\beta$ too large leads to too much added noise and degrades performance in the low SNR regime, while running inference for a very large $N$ is costly and overfits the measurements in the under-sampled regime.

Given a set of $K$ validation noise powers $\{\rho_{i}\}_{i=1\dots K}$, and the resulting validation MSE metrics $n_i(N, \beta)$, we propose selecting the best hyper-parameters by maximizing:
\begin{equation}
    N^\star, \beta^\star = \argmax_{N, \beta} \frac{1}{K} \sum_{i=1}^K \log \rho_i - \log n_i(N, \beta).
    \label{eq:hyper_tune}
\end{equation}

This criterion represents the average inverse estimation error in log-scale across a range of SNR values. Since optimal recovery algorithms for sparse signals exhibit a linear relationship between the noise power and recovery error as long as the number of measurements is above a threshold and the total signal power is normalized \cite{carrillo2016robust}, the above metric is constant across all noise values $\rho_i$. This serves as a principled heuristic for choosing hyper-parameters that perform well across the entire range of values, since it normalizes the residual error by the noise power and promotes low errors at high SNR values, favouring solutions that do not have an error floor.

We search for $\beta$ among the set of values $\{1, 0.1, 0.01, 0.001\}$ and for the stopping point $N$ among $6933$ inference steps. The best pair $(\beta, N)$ is found and stored for each antenna spacing and $\alpha$ value. If the SNR is assumed known, we store a separate pair for each SNR value and we find that, as the SNR increases, the optimal $\beta$ decreases. Intuitively, this corresponds to a higher importance placed on the first two terms, since low SNR scenarios are implicitly regularized by the measurement noise. In the blind (unknown SNR) setting we use \eqref{eq:hyper_tune} to select a single pair of hyper-parameters that maximizes the relative gain in the entire SNR range. The exact values are given in the code.

\begin{figure}[!t]
\centering
\includegraphics[width=0.93\linewidth]{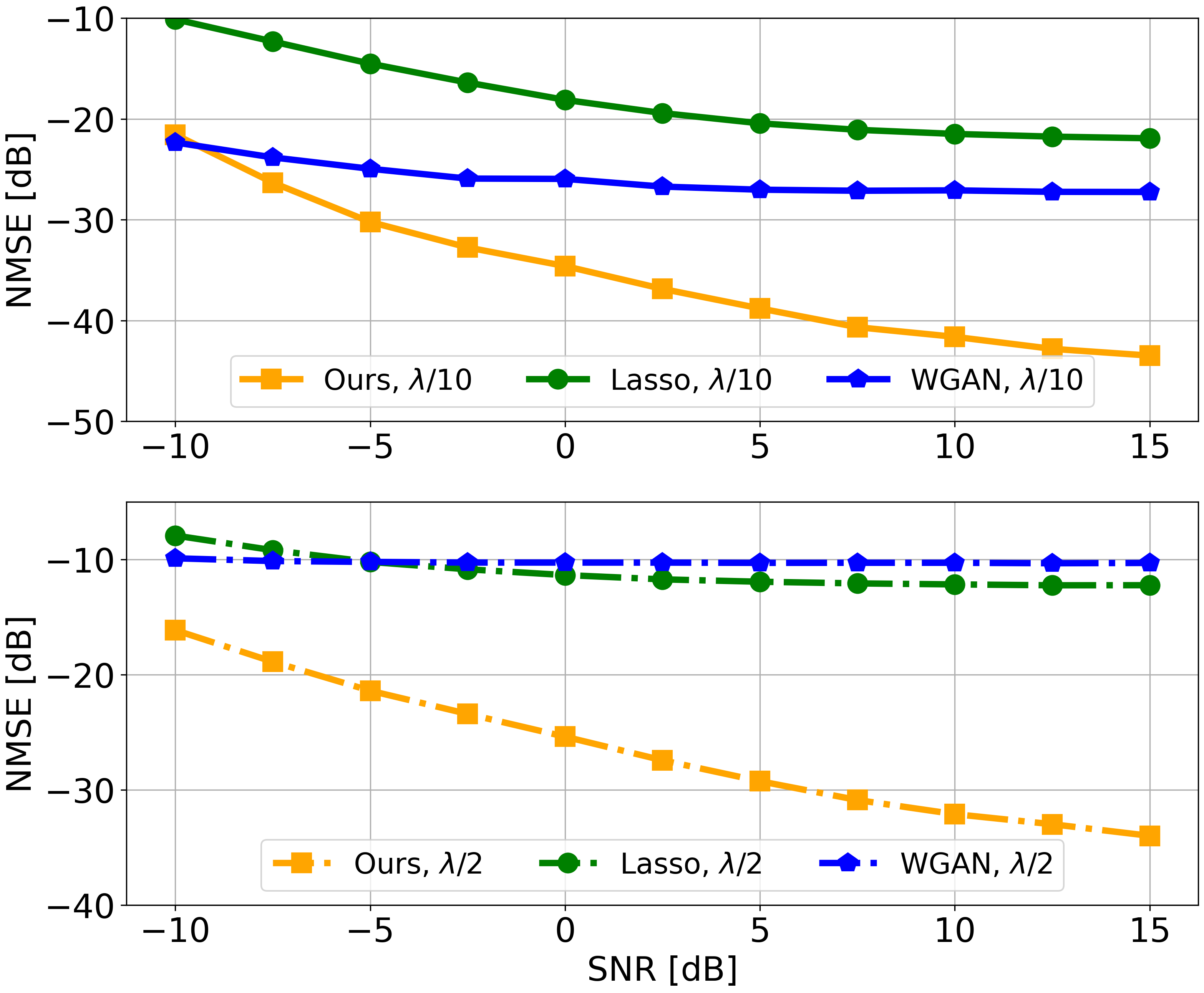}
\caption{In-distribution estimation NMSE on $16\times64$ 5G-NR CDL-D channels at two different values of the antenna spacing. Both deep learning methods (ours and WGAN \cite{balevi2020high}) are trained on CDL-D data at the two wavelengths, and a validation set of $1000$ channel realizations is used to determine the hyper-parameters for all approaches.}
\label{fig:cdl_c_nmse}
\end{figure}

\subsection{In-Distribution Performance}
We compare our results with two baselines. The Lasso baseline \cite{schniter2014channel} uses the $\ell_1$-norm of the 2D-DFT of the channel $\mathbf{h}$ as a regularization term in \eqref{eq:cs}.
We implement this using the SigPy library \cite{sigpy}, a well-optimized Python package designed for inverse problems. For the known SNR setting, we tune a separate $\lambda$ value for each SNR value, otherwise we use the criterion in \eqref{eq:hyper_tune} to find a single value in the blind setting. The optimal values for $\lambda$ vary as a function of $\alpha$, and are generally in the $[0.03, 0.3]$ range.

The second baseline is the method in \cite{balevi2020high} -- which we term Wasserstein GAN (WGAN). We implement and successfully reproduce the authors' results. To further boost the performance of this baseline we tune the size of the latent representation $\mathbf{z}$, and we increase the number of layers by two, as well as the number of hidden channels to match the number of learnable weights. Complete details are available in the code.

\begin{figure*}[!t]
\centering
\includegraphics[width=0.8\linewidth]{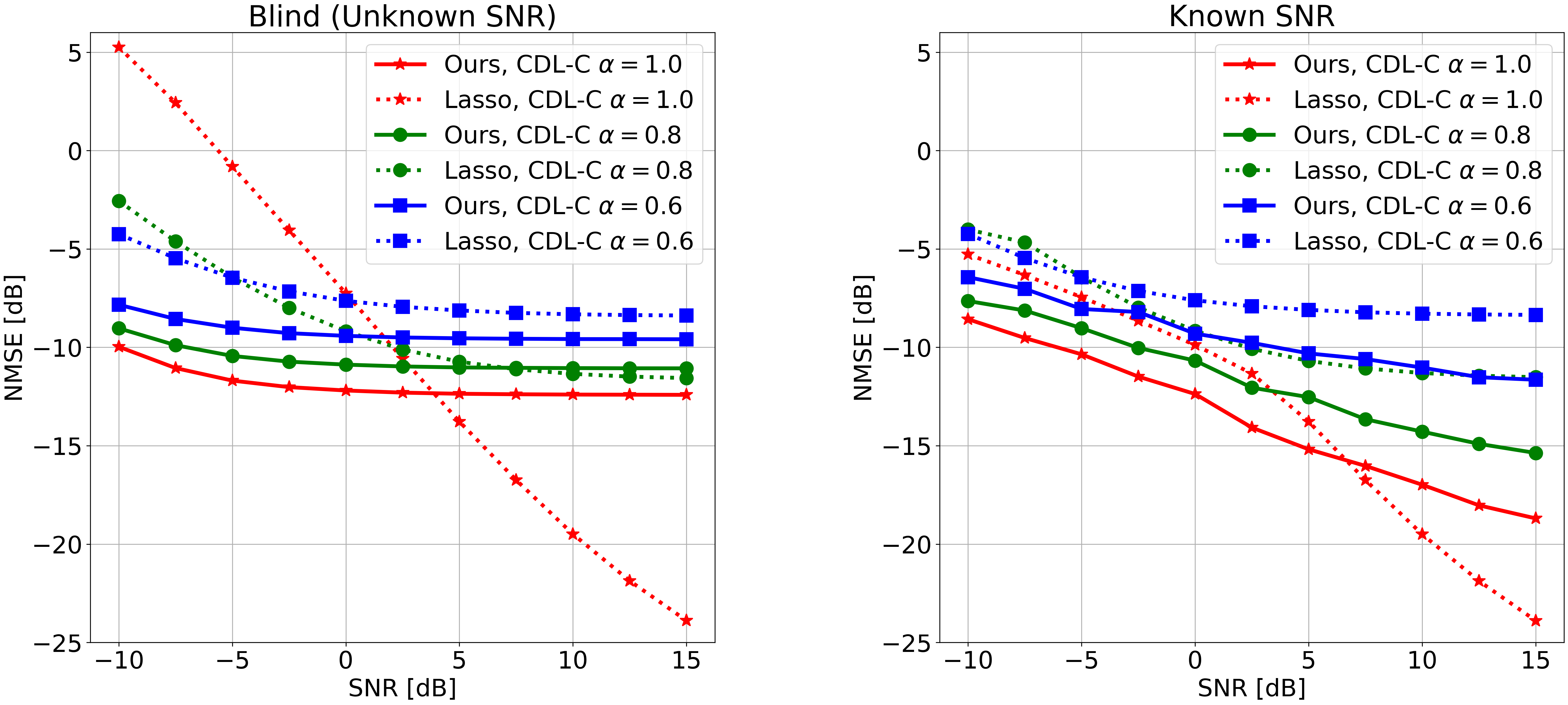}
\caption{Estimation NMSE on $16\times64$ 5G-NR CDL-C channels at $\lambda/2$ antenna spacing in the blind (left plot) and known (right plot) settings, respectively. Our proposed approach is never trained or tuned on any channel realization from the distribution of CDL-C channels and retains its usefulness as a channel estimator and outperforms Lasso when $\alpha < 1$ in both cases.}
\label{fig:cdl_c_nmse_mixed}
\end{figure*}

\begin{figure*}[!t]
\centering
\includegraphics[width=0.95\linewidth]{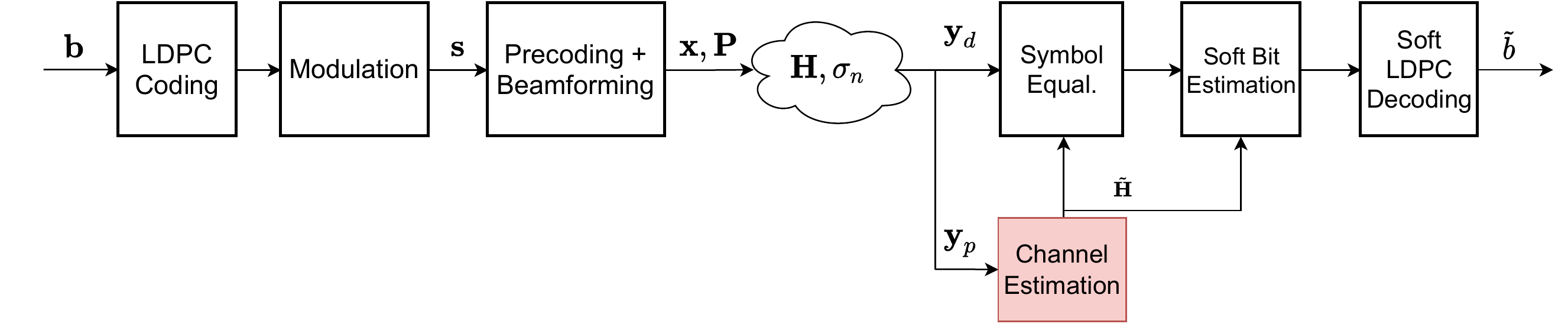}
\caption{Block diagram with the components of the simulated end-to-end scenario. $\mathbf{x}$ and $\mathbf{P}$ represent the data and pilot transmissions, respectively. The red block marks channel estimation, where our proposed approach is used and further feeds into the symbol equalization and maximum-likelihood soft bit estimation blocks. The coded bit error rate is measured between $\mathbf{b}$ and $\tilde{\mathbf{b}}$. This implementation is part of our released source code.}
\label{fig:e2e_block}
\end{figure*}

Fig.~\ref{fig:cdl_c_nmse} shows the test NMSE on in-distribution CDL-D channels in the blind SNR setting with $\alpha = 0.4$. From the top plot it can be seen that for a very low antenna spacing, the WGAN approach captures some of the structure of the channel, but the performance quickly saturates in the high SNR regime, and the same is true for Lasso. This effect is worsened when the antenna spacing is $\lambda/2$ and there is less structure, indicating that neither baseline uses a suitable prior. In contrast, the proposed approach exhibits almost a linear decay of the NMSE, thus being near-optimal and inline with the theoretical results in \cite{jalal2021robust}, all while never explicitly learning a prior. At the SNR value of $15$ dB the NMSE is more than $12$ dB lower than both baselines, but as results in the sequel show, this gap is reduced in end-to-end systems by channel coding.

\subsection{Robust (Out-of-Distribution) Performance}
\begin{figure}[!t]
\centering
\includegraphics[width=0.95\linewidth]{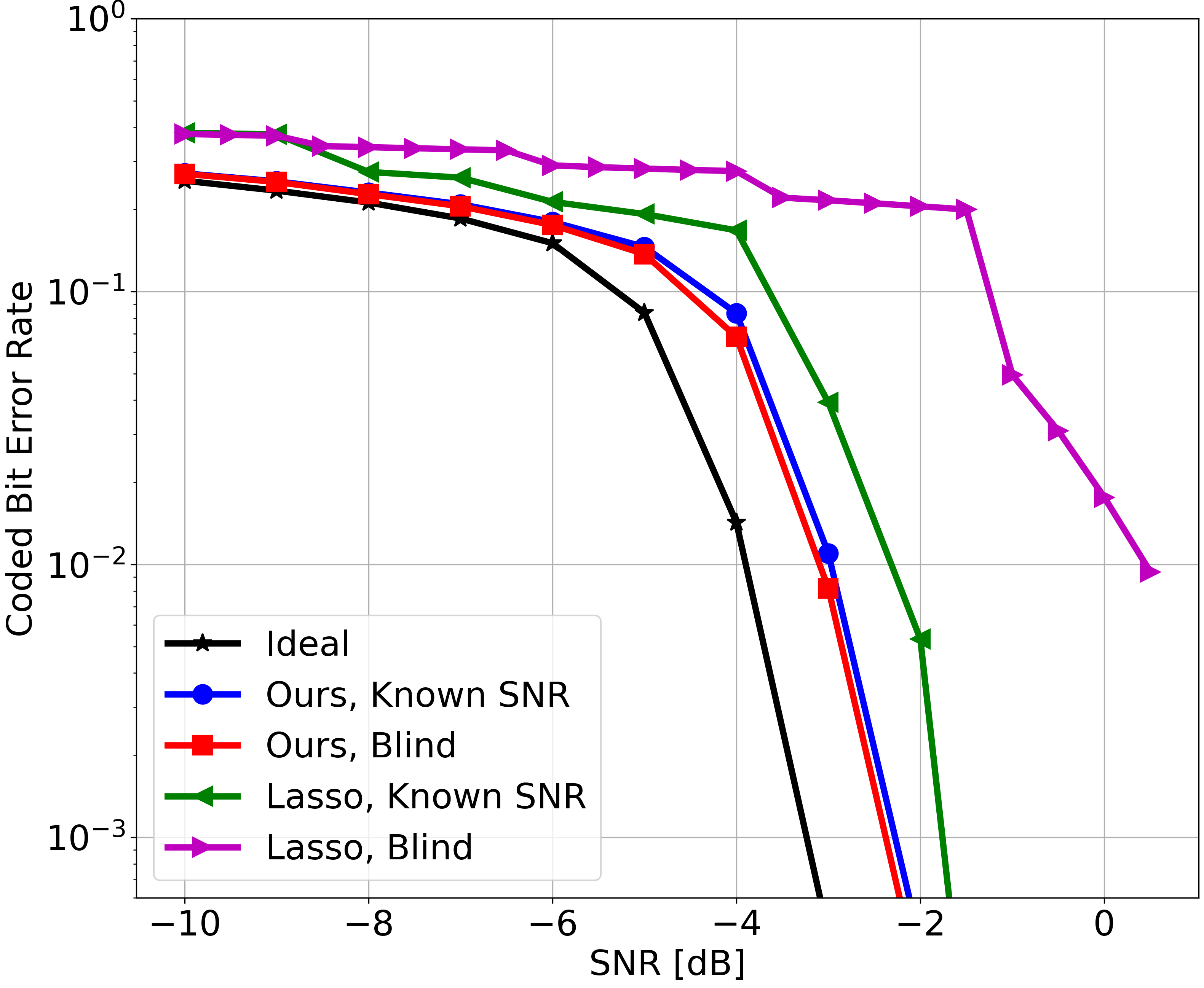}
\caption{End-to-end coded bit error performance in a comprehensive digital communication chain simulation of the proposed method, compared with ideal channel knowledge, in 5G-NR CDL-C channels. Our method is never trained or tuned on any channel realization from the distribution of CDL-C channels and offers excellent and robust end-to-end performance under realistic impairments.}
\label{fig:cdl_c_ber}
\end{figure}
We evaluate the robustness of the proposed approach by using a model trained only with line of sight channels to estimate non-line of sight channels. Compared to the CDL-D channels used for training, CDL-C channels are significantly less sparse due to being more scattered in the delay domain. We omit WGAN performance from these results, since CDL-C channels are not in the range of the trained generative model and the high estimation NMSE makes it unusable out-of-distribution. Fig.~\ref{fig:cdl_c_nmse_mixed} shows test performance of our proposed approach and Lasso when tested on CDL-C channels at antenna spacing $\lambda/2$, in two settings: unknown (blind) and known SNR, respectively. In both cases, our method outperforms Lasso in the low SNR and few measurements regime, however performance still plateaus in the unknown SNR regime.

Finally, we evaluate coded end-to-end performance in a proof-of-concept system shown in Fig.~\ref{fig:e2e_block} in order to evaluate the practical impact of different channel estimation algorithms. The transmitter uses LDPC encoding with a rate of $1/2$ and codeword size $648$ bits, followed by a four stream QPSK modulation and random Gaussian precoding with unit gain to $64$ antennas. The receives estimates the channels from random QPSK pilot transmissions $\mathbf{P}$ with a transmit power boosted by $20$ dB relative to the data signal power, performs symbol equalization, followed by soft bit estimation and LDPC decoding. We measure system performance in coded bit error rate.

Fig.~\ref{fig:cdl_c_ber} shows end-to-end performance of our method compared to Lasso. Note that this experiment is using CDL-C channels with $\alpha=1$, that are completely out-of-distribution for the proposed approach. We use $\alpha=1$ to demonstrate the improved performance in the low SNR regime. We can see that, in the case of blind SNR, Lasso is highly sub-optimal up to a SNR of $0$ dB, exactly where the channel coding cannot correct all errors in the transmission, thus leading to a $4$ dB loss in end-to-end performance. Thus, even though Lasso has a lower NMSE at high SNR values, the loss at low SNR is a determining factor that drives the system performance. In contrast, the proposed approach is robust to unknown SNR and exhibits a $0.5$ dB loss against ideally known channels. Even in the known SNR regime, our method still outperforms Lasso by $0.5$ dB, even though no CDL-C channels were used for training or fine-tuning either approach. Finally, we note that the performance is slightly better for our method in the unknown SNR case: this is due to the fact that the inference hyper-parameters are only tuned on CDL-D channels, but tested on CDL-C, leading to a mismatch in terms of performance.

\section{Discussion and Conclusion}
In this paper, we have introduced a score-based model approach for MIMO wireless channel estimation. Our method performs posterior sampling using a pretrained score-based model and includes a careful tuning of the inference hyper-parameters. When tested on out-of-distribution channels, the proposed approach retains its performance and surpasses the Lasso algorithm in an end-to-end setting.

One current limitation of the proposed approach is its expensive inference time. Posterior sampling for a single channel of size $16\times 64$ takes approximately one minute on a graphical processing unit. On the other hand, Lasso is considerably faster, taking as little as $3$ seconds per channel. However, both approaches are still too slow for real-world deployment, where channels are re-estimated as fast as every millisecond. Speeding up inference with score-based models is an active area of research, with recent work showing speed-ups of up to three orders of magnitude \cite{anonymous2022progressive}. This is a promising future research direction for channel estimation, given the latency-sensitive nature of the task. Finally, in this paper we have only trained and tested score-based models on channels of fixed size. Due to the fully-convolutional nature of our model, a future promising direction is to train on a mixture of sizes along with antenna spacings and fading distributions, to obtain a near-universal channel estimation approach.

\bibliographystyle{IEEEtran}
\bibliography{myBib}

\end{document}